\documentclass [a4paper,fleqn]{article}
\usepackage{graphicx}
\usepackage {subfigure,epsfig}

\usepackage {amsmath} \usepackage{amssymb} \usepackage{cite} \usepackage{amsthm}
\numberwithin{equation}{section} \numberwithin{table}{section} \mathindent=0pt
\theoremstyle{plain}

\numberwithin{theorem}{section}

\begin{document}

\title{\textbf{Relations for zeros of special
polynomials associated to the Painlev\'{e} equations}}

\author{Nikolai A. Kudryashov, Maria V. Demina}

\date{Department of Applied Mathematics\\
Moscow Engineering and Physics Institute\\ (State University)\\
31 Kashirskoe Shosse, 115409, Moscow, \\ Russian Federation}
\maketitle

\begin{abstract}

A method for finding relations for the roots of polynomials is
presented. Our approach allows us to get a number of relations for
the zeros of the classical polynomials and for the roots of special
polynomials associated with rational solutions of the Painlev\'{e}
equations. We apply the method to obtain the relations for the zeros
of several polynomials. They are: the Laguerre polynomials, the
Yablonskii - Vorob'ev polynomials, the Umemura polynomials, the
Ohyama polynomials, the generalized Okamoto polynomials, and the
generalized Hermite polynomials. All the relations found can be
considered as analogues of generalized Stieltjes relations.

\end{abstract}



\section{Introduction}

In 1885 Stieltjes \cite{Stieltjes01, Stieltjes02} suggested the
following relations for the roots of the Hermite polynomials
\cite{Szego01, Veselov01}

\begin{equation}\begin{gathered}
\label{ku0.1} \sum_{k\neq
j}^{n}\left(z_{j}-z_{k}\right)^{-1}-z_{j}=0, \,\,\,\, (j=1,...,n)
\end{gathered}
\end{equation}

These relations can be considered as the equilibrium configurations
of $n$ particles on the line interecting pairwise with a repulse
logarithmic potential in the harmonic field \cite{Veselov01,
Calogero01, Ahmed01}.

In this paper we introduce a new approach for finding relations like
\eqref{ku0.1} satisfied by the roots of polynomials, which can be
generated by a differential equation. We demonstrate our approach to
get relations for the roots of the Laguerre polynomials and for the
roots of special polynomials associated with rational solutions of
the second, the third, and the fourth Painlev\'{e} equations.

The six Painlev\'{e} equations $(P_1-P_6)$ were picked out by
Painlev\'{e} and his student Gambier from a certain class of second
- order differential equations as equations whose general solutions
on the one hand, do not have movable critical points and on the
other hand, cannot be expressed through known elementary or special
functions. In other words they define new transcendental functions.
However, the equations $P_2-P_6$ possess hierarchies of rational and
algebraic solutions at certain values of the parameters. It turned
out that such solutions could be described using families of special
polynomials. The history of this question is as follows.

A. I. Yablonskii and A. P. Vorob'ev were the first who expressed the
rational solutions of $P_2$ via the logarithmic derivative of the
polynomials, which now go under the name of the Yablonskii --
Vorob'ev polynomials \cite{Yablonskii01, Vorob'ev01}. Later K.
Okamoto suggested special polynomials for certain rational solutions
of $P_4$ \cite{Okamoto01}. H. Umemura derived analogous polynomials
for some rational and algebraic solutions of $P_3$  and $P_5$
\cite{Umemura01}. All these polynomials possess a number of
interesting properties. For example, they can be expressed in terms
of Schur polynomials. Besides that the polynomials arise as the so
-- called tau-functions and satisfy recurrence relations of Toda
type. Recently these polynomials have been intensively studied
\cite{Clarkson01, Clarkson02, Clarkson03}.

Apart from rational and algebraic solutions $P_2-P_6$ possess one --
parameter families of solutions expressible in terms of the
classical special functions. In particular cases these special
function solutions are reduced to classical orthogonal polynomials
and, consequently give rational solutions of the Painlev\'{e}
equation in question. For $P_4$ these polynomials are Hermite
polynomials, for $P_3$ and $P_5$ Laguerre polynomials and for $P_6$
Jacobi polynomials.

Not long ago P. A. Clarkson and E. L. Mansfield suggested special
polynomials for the equations of the $P_2$ hierarchy
\cite{Clarkson04}. Also they studied the location of their roots in
the complex plane and showed that the roots have a very regular
structure.

The textbook \cite{Iwasaki01} characterizes the Painlev\'{e}
equations as "the most important nonlinear ordinary differential
equations" and states that "many specialists believe that during the
twenty-first century the Painlev\'{e} equations will become new
members of the community of special functions" \cite{Amdeberhan01}.

In this paper we study the relations for the zeros of the following
equations

\begin{equation}\begin{gathered}
\label{ku0.2}yy_{{{zzzz}}}-4\,y_{{z}}y_{{{zzz}}}+3\,{y_{{{
zz}}}}^{2}-z \left( yy_{{{zz}}}-{y_{{z}}}^{2} \right) -yy_{{z}}=0,
\end{gathered}
\end{equation}

\begin{equation}\begin{gathered}
\label{ku0.3}z^2\,(y\,y_{zzzz}-4\,y_z\,y_{zzz}+3\,y_{zz}^2)+2\,z\,(y\,y_{zzz}-
y_{z}\,y_{zz})\\
-4\,z\,(z+\mu)(\,y\,y_{zz}-y_{z}^2)-2\,y\,y_{zz}+4\,\mu\,y\,y_{z}-2\,n\,(n+1)\,y^2=0,
 \end{gathered}
\end{equation}

\begin{equation}\begin{gathered}
\label{ku0.4}z^3\,(y\,y_{zzzz}-4\,y_z\,y_{zzz}+3\,y_{zz}^2)-6\,z^2\,(y\,y_{zzz}-
y_{z}\,y_{zz})\\
-12\,z\,(z^4-3\,n-1)(\,y\,y_{zz}-y_{z}^2)-9\,z\,(y\,y_{zz}+y_{z}^2)\\
+3(12\,z^4-16\,n\,z^2+12\,n+7)y\,y_{z}\\
-24\,n\,z\,((n+3)\,z^2-3\,n-1)\,y^2=0,
 \end{gathered}
\end{equation}

\begin{equation}\begin{gathered}
\label{ku0.5}y\,y_{zzzz}-4\,y_z\,y_{zzz}+3\,y_{zz}^2
+\frac{4}{3}\,z^2\,(y\,y_{zz}-y_{z}^2)
+4\,z\,y\,y_z\\
-\frac{8}{3}\,(m^2+n^2+m\,n-m-n)\,y^2=0,
 \end{gathered}
\end{equation}

\begin{equation}\begin{gathered}
\label{ku0.6}y\,y_{zzzz}-4\,y_z\,y_{zzz}+3\,y_{zz}^2-4(z^2+2n-2m)\,(y\,y_{zz}-y_{z}^2)\\
+ 4\,z\,y\,y_z-8\,m\,n\,y^2=0.
 \end{gathered}
\end{equation}

It is known \cite{Clarkson01} that particular solutions of equation
\eqref{ku0.2} are the Yablonskii - Vorob'ev polynomials. Rational
solutions of the second Painlev\'{e} equation can be expressed as
the logarithmic derivative of these polynomials. In this work using
our approach we obtain some new relations for the zeros of these
polynomials. These relations explain a very regular location of the
zeros in the complex plain.

The equations \eqref{ku0.3} and \eqref{ku0.4} in their turn possess
polynomial solutions, i.e. the Umemura polynomials and the Ohyama
polynomials \cite{Clarkson02}, accordingly. Rational solutions of
the third Painlev\'{e} equation can be obtained with the help of
these polynomials. The general form of these polynomials is not
known at present but using the method of our recent works
\cite{Kudryashov01, Demina01} one can solve this problem and can get
the general form for these polynomials. However here we present new
additional correlations for the zeros of these polynomials.

The equations \eqref{ku0.5} and \eqref{ku0.6} also admit polynomial
solutions, which are the generalized Okamoto polynomials and the
generalized Hermite polynomials, respectively \cite{Clarkson03}.
Rational solutions of the fourth Painlev\'{e} equation can be
expressed via logarithmic derivative of these polynomials. The
general form of these polynomials is not known as well. In this work
new correlations for the zeros of these polynomials are obtained.

This paper is organized as follows. In section 2 we present the
method of finding the relations for the roots of the polynomials. In
this section we also apply our approach on the example to have the
relations for the roots of the Laguerre polynomials. Sections 2, 3,
4 and 5 are devoted to finding the relations for the zeros of the
Yablonskii - Vorob'ev polynomials, of the Umemura polynomials, of
the Ohyama polynomials, of the generalized Okamoto polynomials and
of the generalized Hermite polynomials. All these polynomials are
associated with rational solutions of the second, the third and the
fourth Painlev\'{e} equations.

\section{Method applied}

Let us assume that we have an n-linear (linear, bilinear etc)
equation of order $M$
\begin{equation}\begin{gathered}
\label{2.0a}E_1\left(y,y_z,y_{zz},...,z\right)=0.
\end{gathered}\end{equation}
In addition suppose that this equation admits polynomial solution
$y(z)=q(z)$ of degree $p$. The transformation
\begin{equation}\begin{gathered}
\label{2.0b} w(z)=\frac{y_z}{y}
\end{gathered}\end{equation}
maps solutions of the equation \eqref{2.0a} into solutions of the
nonlinear equation
\begin{equation}\begin{gathered}
\label{2.0c}E_2\left(w,w_z,w_{zz},...,z\right)=0.
\end{gathered}\end{equation}
Note that the degree of \eqref{2.0c} is $M-1$. Under such
transformation the polynomial $q(z)$ becomes the rational function
\begin{equation}
\begin{gathered}
\label{2.6}w(z)=\sum_{k=1}^{\tilde{p}}\frac{e_k}{z-z_{k}},
\end{gathered}
\end{equation}
where $z_j$ is a root of $q(z)$ of degree $e_j$ and
\begin{equation}
\begin{gathered}
\label{2.6a}p=\sum_{j=1}^{\tilde{p}}e_j,
\end{gathered}
\end{equation}
i.e. $\tilde{p}$ is the number of distinct roots. The rational
function \eqref{2.6} can be expanded in a neighborhood of a pole
$z_{j}$ and this gives
\begin{equation}\label{2.7}
\begin{gathered}
w(z)=\frac{e_j}{z-z_{j}}-\sum_{m=0}^{\infty} \left( \sum_{k\neq
i}\frac{e_k}{(z_{k}-z_{j})^{m+1}} \right)
(z-z_{j})^{m},\\
\\
0<\mid z-z_{j}\mid< \min_{k \neq j} \mid z_{k}-z_{j}\mid
\end{gathered}
\end{equation}

At the same time we can find expansions of solution $w(z)$ near
movable singular point $z_0$ using the Painlev\'{e} test
\cite{Ablowitz01, Conte01, Kudryashov02, Mugan01} or methods of
power geometry\cite{Bruno01, Bruno02, Bruno03, Kudryashov03,
Demina02}. The result of calculations for the rational function
$w(z)$ takes the form
\begin{equation}\label{2.8}
\begin{gathered}
w(z)=\sum_{m=0}^{\infty}c_{m}^{}(z-z_0)^{m-1},
\end{gathered}
\end{equation}

Assuming $z_0=z_{j}$, $c_0=e_j$ and comparing this series with
\eqref{2.7} yields

\begin{equation}\label{2.9}
\begin{gathered}
\forall j: \quad \sum_{k\neq
i}\frac{e_k}{(z_{k}-z_{j})^{m+1}}=-c_{m}^{},\quad m>0.
\end{gathered}
\end{equation}

We are going to use correlations \eqref{2.9} for finding some
relations of the zeros of the above mentioned polynomials, which are
associated with the rational solutions of the second, the third, and
the fourth Painlev\'{e} equations.

Let us demonstrate our approach by the example of the Laguerre
polynomials $L_n^{(\alpha)}(z)$. They satisfy the following
second-order linear differential equation
\begin{equation}\begin{gathered}
\label{2.10}z\,y_{zz}+(\alpha+1-z)\,y_{z}+n\,y=0.
\end{gathered}
\end{equation}
Substituting $y_z=w(z)\,y(z)$, $y_{zz}=(w_z+w^2)y$ into \eqref{2.10}
and dividing the result by $y(z)$ we obtain the first-order
nonlinear differential of the form
\begin{equation}\begin{gathered}
\label{2.11}z\,w_z+z\,w^2+\left(\alpha+1-z\right)w+n=0.
\end{gathered}\end{equation}

Taking the Painlev\'{e} methods into account we have the power
expansions for rational solution of equation \eqref{2.11} near
singular point $z=z_0$ $(z_0\neq0)$ in the form

\begin{equation}\begin{gathered}
\label{2.12}w(z)=\frac{1}{z-{z_0}}-\frac12\,{\frac {
\alpha+1-{z_0}}{{z_0}}}+\frac{1}{{{12\,z_0}}^{2}}\,[\,6\,\alpha+5-2\,{
z_0}-2\,\alpha\,{z_0}+\\
\\
+{{z_0}}^{2}-4\,n{z_0}+{\alpha}^{2}\,] \left( z-{z_0}
\right)-\frac{1}{8{{z_0}}^{3}}\,[\,
4\,\alpha+{\alpha}^{2}+3-\alpha\,{z_0}-2\,n{z_0}-\\
\\
-{z_0 }\,]\left( z-{z_0} \right) ^{2}-{\frac {1}
{720{{z_0}}^{4}}}\,[\,76\,
\alpha\,{z_0}-360\,\alpha-8\,n{z_0}\,{\alpha}^{2}+2\,{{
z_0}}^{2}+\\
\\
+152\,n{z_0}-110\,{\alpha}^{2}-4 \,{\alpha}^{2}{
z_0}+8\,\alpha\,{{z_0}}^{2}+16\,n{{z_0}}^{2}-4
\,{{z_0}}^{3}-251+\\
\\
+{{z_0}}^{4}+{\alpha}^{4}-8\,{{z_0}}^{3}n+6
\,{\alpha}^{2}{{z_0}}^{2}-4\,\alpha\,{{z_0}}^{3}-4\,{\alpha}^{3} {
z_0}+\\
\\
+16\,{n}^{2}{{z_0}}^{2}+16\,\alpha\,{{z_0}}^{2}n+76\,{z_0}\,] \left(
z-{z_0} \right) ^{3}+...
\end{gathered}\end{equation}

Comparing the power expansion \eqref{2.12} with \eqref{2.7} we have
the following relations for the zeros of the Laguerre polynomials in
the case $z_j\neq0$

\begin{equation}\begin{gathered}
\label{2.13}\sum_{k\neq
j}^{\tilde{n}}\frac{e_k}{(z_k-z_j)}-\frac{\alpha+1}{2\,z_j}+\frac{1}{2}=0,
\end{gathered}\end{equation}

\begin{equation}\begin{gathered}
\label{2.14}\sum_{k\neq
j}^{\tilde{n}}\frac{e_k}{(z_k-z_j)^2}+\frac{\alpha^2+6\,\alpha+5}{12\,z_{j}^2}-\frac{1+
\alpha+n}{6\,z_{j}}+\frac{1}{12}=0,
\end{gathered}\end{equation}

\begin{equation}\begin{gathered}
\label{2.15}\sum_{k\neq
j}^{\tilde{n}}\frac{e_k}{(z_k-z_j)^3}-\frac{\alpha^2+4\,\alpha+3}{8\,z_{j}^{3}}+\frac{1+
\alpha+2\,n}{8\,z_{j}^2}=0
\end{gathered}\end{equation}
In this expressions $\tilde{n}$ makes sense of $\tilde{p}$ and
$e_k=1$, $z_k\neq0$. It is known that $z=0$ is the root of
$L_n^{(\alpha)}(z)$ if and only if $\alpha=-l$ $(1\leq l\leq n)$.
Then in this case the asymptotic series in a neighborhood of $z=0$
is
\begin{equation}\begin{gathered}
\label{2.16}w(z)={\frac {l}{z}}-{\frac {n-l}{l+1}}-{\frac { \left(
n-l \right)  \left( n+1 \right) z}{ \left( l+2 \right) \left( l+1
\right) ^{2}}}\left[1+ \frac{( 2\,n+1-l  )z}{ \left( l+3 \right)
 \left( l+1\right)}\right.\\
\left.+ \frac{( 5\,{n}^{2}
l+11\,{n}^{2}+11\,n-5\,n{l}^{2}-6\,nl+2-6\,l+{ l}^{3}-{l}^{2})
{z}^{2}}{ ( l+3 )
 ( l+4 ) ( l+2 )  ( l+1 ) ^{2}}+o(z^2) \right]
\end{gathered}\end{equation}

There are no arbitrary constants in this expansion as the critical
number $k$ is $k=-l-1<-1$. Without loss of generality let us suppose
that $z=0$ is the first in the set of roots. Comparing the series
\eqref{2.16} with \eqref{2.7} we see that the order of the root
$z=0$ is $l$ and
\begin{equation}\begin{gathered}
\label{2.17}\sum_{k=2}^{\tilde{n}}\frac{1}{z_k}-{\frac
{n-l}{l+1}}=0,
\end{gathered}\end{equation}

\begin{equation}\begin{gathered}
\label{2.18}\sum_{k=2}^{\tilde{n}}\frac{1}{z_k^2}-{\frac { \left(
n-l \right)  \left( n+1 \right) }{ \left( l+2 \right)
 \left( l+1 \right) ^{2}}}=0,
\end{gathered}\end{equation}

\begin{equation}\begin{gathered}
\label{2.19}\sum_{k=2}^{\tilde{n}}\frac{1}{z_k^3}-{\frac { \left(
n-l \right)  \left( n+1 \right) ( 2\,n+1-l  )}{ \left( l+2
\right)(l+3)
 \left( l+1 \right) ^{3}}}=0
\end{gathered}\end{equation}

We would like to note that in such a way one can find analogous
relations for some other polynomials.

\section{Relations for the zeros of the Yablonskii - Vorob'ev polynomials}

The transformation \eqref{2.0b} maps solutions of the equation
\eqref{ku0.2} into solutions of the third-order nonlinear
differential equation
\begin{equation}\begin{gathered}
\label{3.0}w_{{{zzz}}}+6\,{w_{{z}}}^{2}-zw_{{z}}-w=0.
\end{gathered}\end{equation}
The Laurent expansion of its solution $w(z)$ in a neighborhood of a
movable pole $z_0$ is the following
\begin{equation}\begin{gathered}
\label{3.1}w(z)=\frac{1}{(z-z_0)}+{a_1}+\frac{{z_0}}{12}\,\left(
z-z_{{0}} \right)-\left( {\frac
{{{z_0}}^{2}}{720}}\,+\frac{{a_1}}{30}\right)\left(
z-z_{{0}} \right) ^{3} -\\
\\
-{\frac {{z_0}}{144}}\,\, \left( z-z_{{0}} \right) ^{4}+{a_6}\,
\left( z-z_{{0}} \right) ^{5}- \left({\frac {{{z_0}}^{2}}{8640}}\,+{
\frac {{a_1}}{360}}\, \right) \left( z-z_{{0}} \right) ^{6}+...
\end{gathered}\end{equation}
The  Yablonskii - Vorob'ev polynomials $Q_n(z)$ are monic
polynomials of degree $n(n+1)/2$ with simple roots. Then from power
expansion \eqref{3.1} we obtain the following relations for the
zeros of $Q_n(z)$
\begin{equation}\begin{gathered}
\label{3.2}\sum_{k\neq j}^{n(n+1)/2}\frac{1}{(z_k-z_j)}+a_1=0,
\end{gathered}\end{equation}

\begin{equation}\begin{gathered}
\label{3.3}\sum_{k\neq
j}^{n(n+1)/2}\frac{1}{(z_k-z_j)^2}+\frac{z_j}{12}\,=0,
\end{gathered}\end{equation}

\begin{equation}\begin{gathered}
\label{3.3a}\sum_{k\neq j}^{n(n+1)/2}\frac{1}{(z_k-z_j)^3}=0,
\end{gathered}\end{equation}

\begin{equation}\begin{gathered}
\label{3.4}\sum_{k\neq
j}^{n(n+1)/2}\frac{1}{(z_k-z_j)^4}+\frac{{1}}{30}\,\sum_{k\neq
j}^{n(n+1)/2}\frac{1}{(z_k-z_j)}-{\frac {{{z_j}}^{2}}{720}}\,=0,
\end{gathered}\end{equation}

\begin{equation}\begin{gathered}
\label{3.5}\sum_{k\neq
j}^{n(n+1)/2}\frac{1}{(z_k-z_j)^5}-\frac{z_j}{144}\,=0,
\end{gathered}\end{equation}

\begin{equation}\begin{gathered}
\label{3.6}\sum_{k\neq j}^{n(n+1)/2}\frac{1}{(z_k-z_j)^6}+a_6=0,
\end{gathered}\end{equation}

\begin{equation}\begin{gathered}
\label{3.7}\sum_{k\neq j}^{n(n+1)/2}\frac{1}{(z_k-z_j)^7}+{ \frac
{1}{360}}\,\sum_{k\neq j}^{n(n+1)/2}\frac{1}{(z_k-z_j)}-{\frac
{{{z_j}}^{2}}{8640}}\,=0
\end{gathered}\end{equation}

In this expressions the coefficients $a_1$ and $a_6$ depend on $n$
and $z_j$. Further the coefficient $a_6$ can be calculated if we
note that the equation \eqref{3.0} possesses the first integral
\begin{equation}\begin{gathered}
\label{3.8}w_{zz}^2-\frac12w_{zz}+4w_z^3-zw_z^2-2ww_z+\frac12zw+C=0.
\end{gathered}\end{equation}
The latter equation has rational solutions in the form
$w(z;n)=d_z\ln(Q_n(z))$ if $C=-n(n+1)/4$. The Laurent expansion of
$w(z;n)=d_z\ln(Q_n(z))$ coincide with \eqref{3.1} accurate to the
value of $a_6$, which is no longer arbitrary
\begin{equation}\begin{gathered}
\label{3.9}a_{{6}}={\frac {1}{30240}}\,{{\it z_j}}^{3}-{\frac
{1}{280}}-{\frac {1} {420}}\,{\it z_j}\,a_{{1}}+{\frac
{1}{560}}\,n+{\frac {1}{560}}\,{n}^{2 }.
\end{gathered}\end{equation}

We do not present here the Yablonskii - Vorob'ev polynomials. They
were obtained before\cite{Yablonskii01, Vorob'ev01, Clarkson01,
Demina01}. Let us give here only the polynomial $Q_7(z)$. That is

\begin{equation}\begin{gathered}
\label{3.8}Q_7 \left( z \right) =[
{z}^{27}+504\,{z}^{24}+75600\,{z}^{21}+
5174400\,{z}^{18}+62092800\,{z}^{15}+\\
+13039488000\,{z}^{12}-
828731904000\,{z}^{9}-\\
-49723914240000\,{z}^{6}-3093932441600000]z.
\end{gathered}\end{equation}

\begin{figure}[h]
\centerline{\epsfig{file=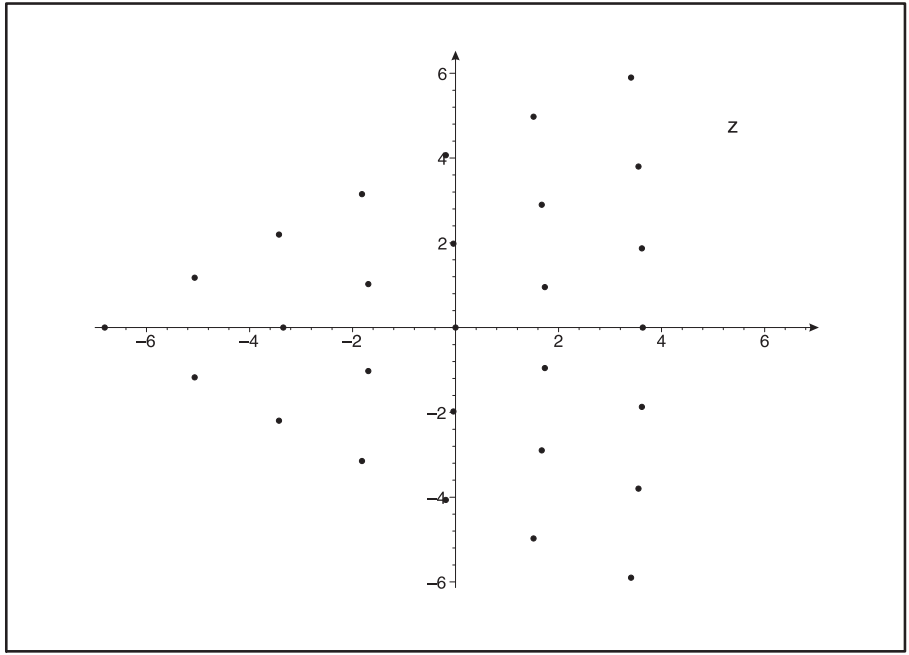,angle=-90,width=120mm}}
\caption{Configuration of zeros for the Yablonskii - Vorob'ev
polynomials $Q_7(z)$}\label{Fig1:z_post}
\end{figure}

The configuration of the zeros for the Yablonskii - Vorob'ev
polynomial $Q_7(z)$ is illustrated in the Figure 1.

\section{Relations for the zeros of the Umemura polynomials}

Using the formula $y_z=w(z)\,y(z)$ we can transform equation
\eqref{ku0.3} to the following third-order nonlinear equation
\begin{equation}\begin{gathered}
\label{4.0}{z}^{2}w_{{{zzz}}}+6\,{z}^{2}{w_{{z}}}^{2}+2\,z\,w_{{{
zz}}}+4\,z
ww_{{z}}-4\,{z}^{2}w_{{z}}-4\,z\,w_{{z}}\mu-2\,w_{{z}}-\\
\\-2\,{w}^{2}+4\,\mu \,w-2\,{n}^{2}-2\,n=0.
\end{gathered}\end{equation}
With the help of the Painlev\'{e} methods \cite{Ablowitz01} we find
the power expansion of its solutions $w(z)$ near movable singularity
$z_0\neq 0$
\begin{equation}\begin{gathered}
\label{4.1}w(z) =\frac{1}{z-z_0}+a_{{1}}-{ \frac { \left(
a_{{1}}-z_{{0}}-\mu \right) }{3\,z_{{0}}}} \left( z-z_{{0}} \right)
-\frac{1}{45{{z_{{0}}}^{3}}}\,[\,4\,{a_{{1}}
}^{2}z_{{0}}-\\
\\
-2\,{z_{{0}}}^{2}a_{{1}}-8\,\mu\,a_{{1}}z_{{0}}+3\,{n}^{2}z_{{0}}+3\,z_{{0}}-9\,a_
{{1}}+9\,\mu+3\,nz_{{0}}+2\,\mu\,{z_{{0}}}^{2}+\\
\\
+{\mu}^{2}z_{{0}}+{z_{{0 }}}^{3}] \left( z-z_{{0}} \right) ^{3}
+\frac{1}{9{z_{{0}}}^{4}}\,[\,
\mu\,{z_{{0}}}^{2}-{z_{{0}}}^{2}a_{{1}}+2\,{a_{{1}}}^{
2}z_{{0}}-4\,\mu\,a_{{1}}z_{{0}}+\\
\\
+{\mu}^{2}z_{{0}}+3\,\mu+{n}^{2}z_{{0} }+z_{{0}}-3\,{a_1}+nz_{{0}}\,
] \left( z-z_{{0}} \right) ^{4}+a_{{6}} \left( z-z_{{0}} \right)
^{5}+...
\end{gathered}\end{equation}

Making use of the expansion near infinity for $w(z)$ we easily find
the degree $p$ of the Umemura polynomials $p=n(n+1)/2$. Proceeding
in the way stated in section 2 we obtain the following relations for
the zeros of the Umemura polynomials in the case $z_j\neq0$

\begin{equation}\begin{gathered}
\label{4.2}\sum_{k\neq j}^{\tilde{p}}\frac{e_k}{(z_k-z_j)}+a_1=0,
\end{gathered}\end{equation}

\begin{equation}\begin{gathered}
\label{4.3}\sum_{k\neq j}^{\tilde{p}}\frac{e_k}{(z_k-z_j)^2}-{ \frac
{ \left( a_{{1}}-\mu \right) }{3\,z_{{j}}}}+\frac{1}{3}=0,
\end{gathered}\end{equation}

\begin{equation}\begin{gathered}
\label{4.4}\sum_{k\neq j}^{\tilde{p}}\frac{e_k}{(z_k-z_j)^3}=0,
\end{gathered}\end{equation}

\begin{equation}\begin{gathered}
\label{4.5}\sum_{k\neq
j}^{\tilde{p}}\frac{e_k}{(z_k-z_j)^4}-\frac{(\mu-a_1)}{5\,z_{j}^{3}}-
\frac{2\,(\mu-a_1)}{45\,z_j}-\frac{1}{45}-\\
\\
-\frac{(\,4\,{a_{{1}}
}^{2}-8\,\mu\,a_{{1}}+\mu^2+3+3\,n+3\,n^2)}{45\,{{z_{{j}}}^{2}}}\,
=0.
\end{gathered}\end{equation}
Here $e_k=1$ for all $z_k\neq0$. Substituting $a_1$ from \eqref{4.2}
into  \eqref{4.3} we obtain
\begin{equation}\begin{gathered}
\label{4.6}\sum_{k\neq
j}^{n}\frac{1}{(z_k-z_j)^2}+\frac{1}{3\,z_j}\,\sum_{k\neq
j}^{n}\frac{1}{z_k-z_j}+\frac{\mu}{3\,z_j}+\frac{1}{3}=0.
\end{gathered}\end{equation}

The Umemura polynomials $Q_5(z)$ takes the form \cite{Clarkson03}

\begin{equation}\begin{gathered}
\label{4.7}Q_{{5}}(z)={z}^{15}-135\,{z}^{14}+8505\,{z}^{13}-331380\,{z}^{12}+8921745
\,{z}^{11}-\\
-175642425\,{z}^{10}+2609560800\,{z}^{9}-29764778400\,{z}^{8
}+\\
+262512608400\,{z}^{7}-1788313917600\,{z}^{6}+9322366561200\,{z}^{5}-
\\
-36474213006000\,{z}^{4}+103536532638000\,{z}^{3}-200979491328000\,{z}^
{2}+\\
+238097325312000\,z-129459762432000
\end{gathered}\end{equation}

\begin{figure}[h]
\centerline{\epsfig{file=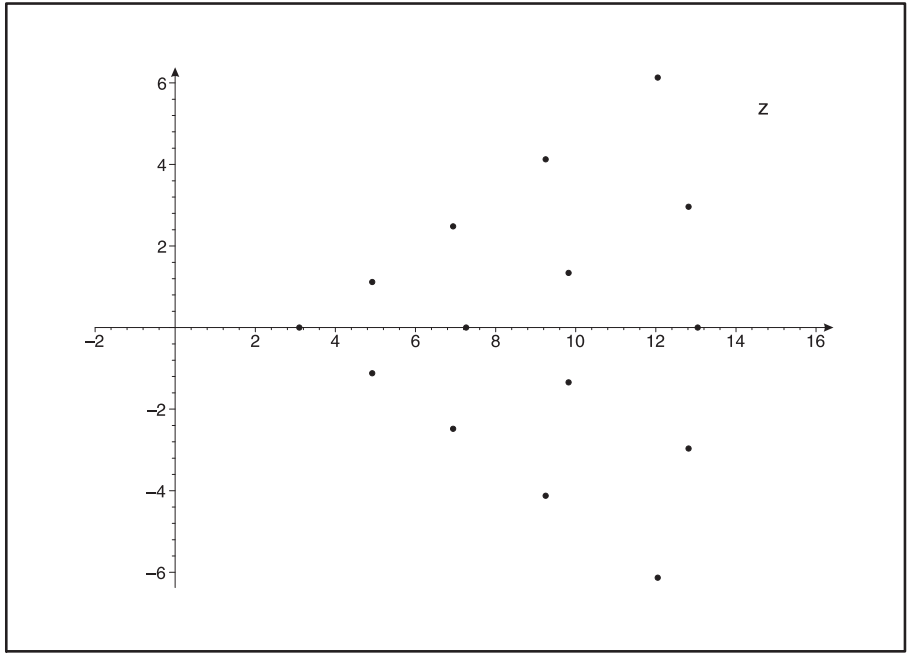,angle=-90,width=120mm}}
\caption{Configuration of zeros for the Umemura polynomial $Q_5(z)$
at $\mu=16$}\label{Fig2:z_post}
\end{figure}

The configuration of the zeros for the Umemura polynomial $Q_5(z)$
is demonstrated in the Figure 2.

\section{Relations for the zeros of the Ohyama polynomials}

Using $y_z=w(z)\,y$ we have from \eqref{ku0.4} the following
third-order differential equation
\begin{equation}\begin{gathered}
\label{5.0}{z}^{3}w_{{{zzz}}}-6\,{z}^{2}w_{{{
zz}}}+6\,{z}^{3}{w_{{z}}}^{2 }+ \left(
-12\,{z}^{2}w-12\,{z}^{5}+36\,nz+3\,z \right) w_{{z}}-\\
-18\,z{ w}^{2}+ \left( 21+36\,{z}^{4}-48\,n{z}^{2}+36\,n \right)
w+24\,nz-\\
-24\, {n}^{2}{z}^{3}-72\,n{z}^{3}+72\,{n}^{2}z=0.
\end{gathered}\end{equation}
The power expansion for the solutions of equation \eqref{5.0} in a
neighborhood of a pole $z=z_0$ $(z_0\neq0)$ is
\begin{equation}\begin{gathered}
\label{5.1}w(z) =
\frac{1}{z-z_0}+a_{{1}}+\frac{1}{{4\,z_{{0}}}^{2}}{ { \left(
4\,a_{{1}}z_{{0}}-7-12\,n+4\,{z_{{0}}}^{4} \right)
}{}} \left( z-z_{{0}} \right)+\\
\\
+\frac{1}{{8\,z_{{0}}}^{3}}{{ \left(21-16\,
a_{{1}}z_{{0}}+36\,n+20\,{z_{{0}}}^{4}-16\,n{z_{{0}}}^{2} \right)
}{}}\left( z-{z_0} \right) ^{2}-\\
\\
-{\frac
{1}{80{{z_{{0}}}^{4}}}}\,[\,64\,{n}^{2}{z_{{0}}}^{4}+64\,{z_{{0}}}^{2}{a_{{1}}}^{2}
-64\,a_{{1}}{z_{{0}}}^{5}+96\,n{z_{{0}}}^{4}+120\,{z_{{0}}}^{4}+144\,{
n}^{2}+\\
\\
+128\,na_{{1}}{z_{{0}}}^{3}+16\,{z_{{0}}}^{8}+301-320\,n{z_{{0}}
}^{2}-192\,na_{{1}}z_{{0}}-192\,{n}^{2}{z_{{0}}}^{2}-\\
\\
-320\,a_{{1}}z_{{0 }} +600\,n\,]\left( z-z_{{0}} \right) ^{3}
+\frac{1}{32{{z_{{0}}}^{5}}}
\,[161-64\,{z_{{0}}}^{6}n+64\,{n}^{2}{z_{{0}}}^{4}+\\
\\
+64\,{z_{{0}}}^{2}{a_{{1}}}^{2}-64\,a_{{1}}{z_{{0}}}^{5}
+96\,n{z_{{0}}}^{4}
+104\,{z_{{0}}}^{4}+144\,{n}^{2}+128\,na_{{1}}{z_{{0}}}^{3}+
\\
\\
+16\,{z_{{0 }}}^{8}-208\,n{z_{{0}}}^{2} -192\,na_{{1}}z_{{0}}
-192\,{n}^{2}{z_{{0}}} ^{2}-208\,a_{{1}}z_{{0}}+\\
\\
+360\,n ] \left( z-z_{{0}} \right) ^{4} +a_{{6}} \left( z-z_{{0}}
\right) ^{5}+...
\end{gathered}\end{equation}
The degree of the Ohyama polynomial can be obtained with the help of
the expansion of $w(z)$ near infinity and equals $ n(n+3)/2$. From
power expansion \eqref{5.2} we obtain the following relations for
the roots of the Ohyama polynomials in the case $z_j\neq0$

\begin{equation}\begin{gathered}
\label{5.2}\sum_{k\neq j}^{\tilde{p}}\frac{e_k}{z_k-z_j}+a_1=0,
\end{gathered}\end{equation}

\begin{equation}\begin{gathered}
\label{5.3}\sum_{k\neq
j}^{\tilde{p}}\frac{e_k}{(z_k-z_j)^2}+z_{j}^{2}+\frac{a_1}{z_j}-\frac{(7+12\,n)}
{4\,z_{j}^{2}}=0,
\end{gathered}\end{equation}

\begin{equation}\begin{gathered}
\label{5.4}\sum_{k\neq
j}^{\tilde{p}}\frac{e_k}{(z_k-z_j)^3}+\frac{5}{2}\,z_{j}-\frac{2\,n}{z_j}-
\frac{2\,a_1}{z_{j}^{2}}+
\frac{(21+36\,n)}{8\,z_{j}^{3}}=0,
\end{gathered}\end{equation}

\begin{equation}\begin{gathered}
\label{5.5}\sum_{k\neq
j}^{\tilde{p}}\frac{e_k}{(z_k-z_j)^4}-\frac{z_{j}^{4}}{5}+
\frac{4\,a_1}{5}\,z_j-\frac{(8\,n^2+12\,n+15)}{10}-\frac{8\,n\,a_1}{5\,z_{j}}-\\
\\
-\frac{(4\,a_{1}^{2}-20\,n-12\,n^2)}{5\,z_{j}^{2}}+\frac{4\,a_1\,(3\,n+4)}{5\,z_{j}^{3}}
-{\frac{(144\,{ n}^{2}+301 +600\,n)}{80{{z_{{j}}}^{4}}}}=0
\end{gathered}\end{equation}

In this expressions $e_k=1$ if $z_k\neq0$. Substituting $a_1$ from
\eqref{5.2} into \eqref{5.3} we have
\begin{equation}\begin{gathered}
\label{5.6}\sum_{k\neq
j}^{\tilde{p}}\frac{e_k}{(z_k-z_j)^2}-\frac{1}{z_j}\,\sum_{k\neq
j}^{\tilde{p}}\frac{e_k}{(z_k-z_j)}+z_{j}^{2}-\frac{{{\left(7+12\,n\right)}}}
{{4\,z_{{j}}}^{2}}=0.
\end{gathered}\end{equation}

The Ohyama polynomial $P_7(z)$ can be written in the form
\cite{Clarkson03}

\begin{equation}\begin{gathered}
\label{5.7}P_7(z)
={z}^{11}[{z}^{24}-56\,{z}^{22}+1470\,{z}^{20}-23800\,{z
}^{18}+263375\,{z}^{16}-\\
-2088240\,{z}^{14}+12105940\,{z}^{12}-51466800
\,{z}^{10}+158533375\,{z}^{8}-\\
-343343000\,{z}^{6}+493643150\,{z}^{4}-
421821400\,{z}^{2}+163788625]
\end{gathered}\end{equation}

\begin{figure}[h]
\centerline{\epsfig{file=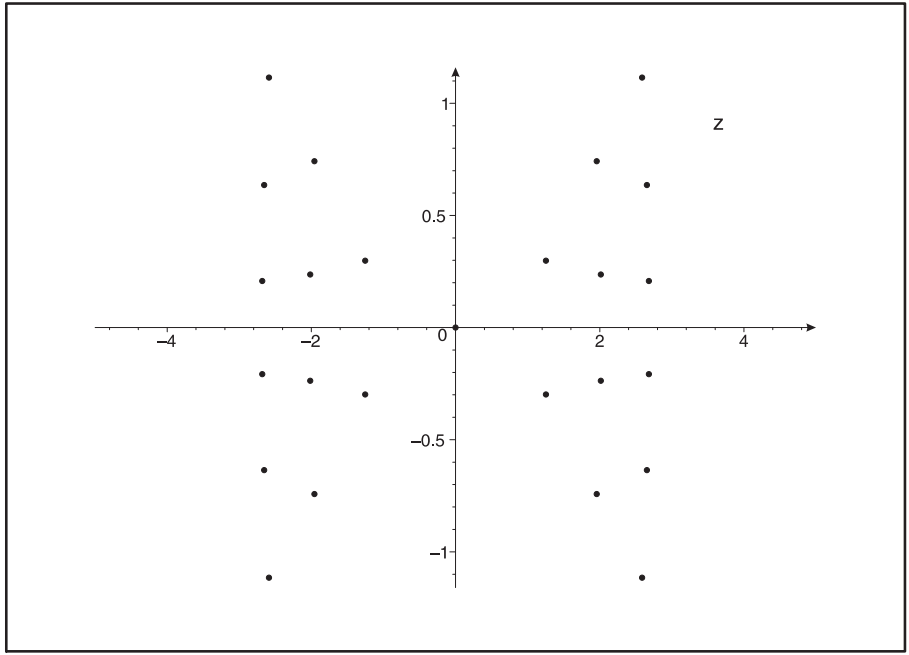,angle=-90,width=120mm}}
\caption{Configuration of zeros for the Ohyama polynomial
$P_7(z)$}\label{Fig3:z_post}
\end{figure}

The configuration of the zeros for the Ohyama polynomial $P_7(z)$ is
presented in the Figure 3.

\section{Relations for the zeros of the generalized Okamoto
polynomials}

Substituting the following expression $y_z=w(z)y$ into \eqref{ku0.5}
sufficient amount of times we get
\begin{equation}\begin{gathered}
\label{6.0}w_{{{zzz}}}+6\,{w_{{z}}}^{2}+\frac43\,{z}^{2}w_{{z}}+4\,z\,w+
\frac83\,\left({m}^{2 }+\,{n}^{2}+\,mn-\,m-\,n\right)=0.
\end{gathered}\end{equation}

The Laurent expansion for solutions of this equation near movable
singularity $z=z_0$ is
\begin{equation}\begin{gathered}
\label{6.1}w(z) =\frac{1}{z-z_0}+a_{{1}}-\frac{{z_{{0 }}}^{2}}{9}\,
\left( z-z_{{0}} \right) +\frac{z_{{0}}}{18}\, \left( z-z_{{0}}
\right) ^{2}+ \\
\\
+\left( {\frac {4}{45}}\,n-{\frac
{4}{45}}\,{m}^{2}+\frac{2z_{{0}}a_{{1}}}{15} \,-{\frac
{4}{45}}\,mn+{\frac {4}{45}}\,m-{\frac {1}{405
}}\,{z_{{0}}}^{4}+{\frac {4}{45}}-{\frac {4}{45}}\,{n}^{2} \right)
\\
\\
\left( z-z_{{0}} \right) ^{3}+ \left( \frac{a_{{1}}}{6}-{\frac
{5}{162}}\,{z_{{0}}}^{3 } \right)  \left( z-z_{{0}} \right)
^{4}+{a_6}\, \left( z-z_{{0}} \right) ^{5}+{\frac
{z_{{0}}}{9720}}\,[\,8\,{z
_{{0}}}^{4}-\\
\\
-432\,z_{{0}}a_{{1}}+288\,mn-288\,n +288\,{m}^{2}-363+288\,{
n}^{2}-288\,m]  \left( z-z_{{0}} \right) ^{6}
\end{gathered}\end{equation}
Then relations for the zeros of the generalized Okamoto polynomials
can be written as
\begin{equation}\begin{gathered}
\label{6.2}\sum_{k\neq j}^{p}\frac{1}{(z_k-z_j)}+a_1=0,
\end{gathered}\end{equation}

\begin{equation}\begin{gathered}
\label{6.3}\sum_{k\neq
j}^{p}\frac{1}{(z_k-z_j)^2}-\frac{z_{j}^2}{9}=0,
\end{gathered}\end{equation}

\begin{equation}\begin{gathered}
\label{6.4}\sum_{k\neq j}^{p}\frac{1}{(z_k-z_j)^3}+\frac{z_j}{18}=0,
\end{gathered}\end{equation}

\begin{equation}\begin{gathered}
\label{6.5}\sum_{k\neq j}^{p}\frac{1}{(z_k-z_j)^4}-{\frac
{{z_{{j}}}^{4}}{405 }}\,\, +\,\frac{2\,a_{{1}}\,z_{{j}}}{15}
+\\
\\
+\frac{4}{45}\,\left(1+m+n-{m}^{2}-m\,n-{n}^{2}\,\right)=0,
\end{gathered}\end{equation}
where the degree $p$ of the generalized Okamoto polynomial
$P_{n,m}(z)$ is $p=m^2+n^2+mn-m-n$.

The generalized Okamoto polynomials $P_{4,4}(z)$ takes the
form\cite{Clarkson02}

\begin{equation}\begin{gathered}
\label{6.6}P_{4,4} \left( z \right) ={z}^{40}-{\frac
{1755}{2}}\,{z}^{36}+{\frac { 4238325}{16}}\,{z}^{32}-{\frac
{310134825}{8}}\,{z}^{28}+\\
+{\frac { 359947121625}{128}}\,{z}^{24}-{\frac
{40989148261425}{256}}\,{z}^{20}+ {\frac
{5340997262498625}{2048}}\,{z}^{16}+\\
+{\frac {299607997122354375} {2048}}\,{z}^{12}-{\frac
{35054135663315461875}{65536}}\,{z}^{8}+\\
+{ \frac {849388671841874653125}{131072}}\,{z}^{4}+{\frac {
1528899609315374375625}{1048576}}
\end{gathered}\end{equation}

\begin{figure}[h]
\centerline{\epsfig{file=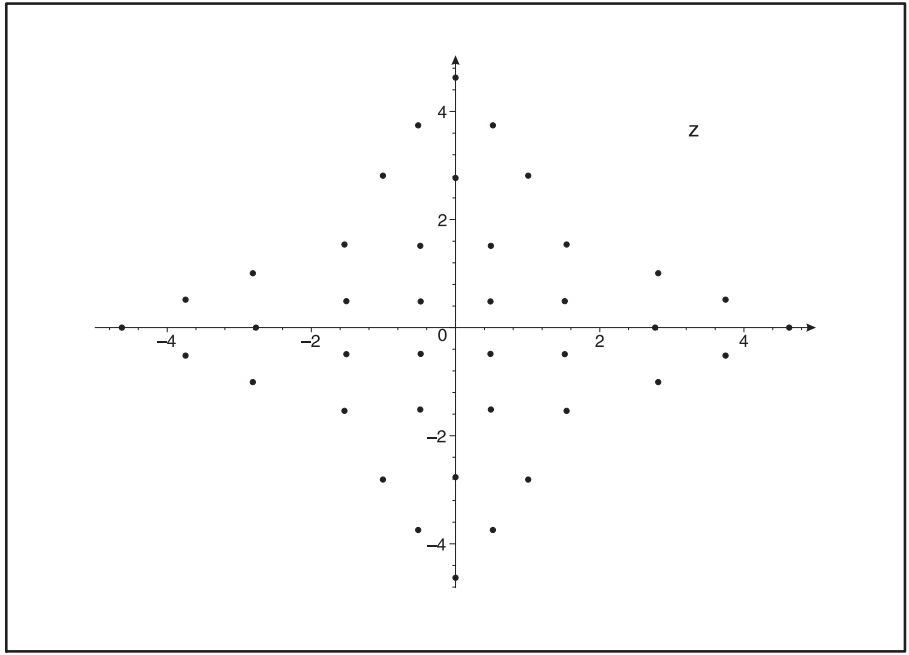,angle=-90, width=120mm}}
\caption{Configuration of zeros for the Okamoto polynomial
$P_{4,4}(z)$}\label{Fig4:z_post}
\end{figure}

The configuration of zeros for the Okamoto polynomial $P_{4,4}(z)$
are demonstrated in the Figure 4.

\section{Relations for the zeros of the generalized
Hermite polynomials}

Applying the transformation $y_z=w(z)y$ we get from \eqref{ku0.6}
the following third-order differential equation
\begin{equation}\begin{gathered}
\label{7.0}w_{{{zzz}}}+6\,{w_{{z}}}^{2}-4\,w_{{z}}{z}^{2}-8\,w_{{z}}
\left(n-m\right)+4\,z\,w-8\,m\,n=0.
\end{gathered}\end{equation}
The power expansion for its solutions near the point $z=z_0$ is
\begin{equation}\begin{gathered}
\label{7.1}w(z) =\frac{1}{( z-{z_0})}+{a_1}+\frac13\left( {{
z_0}}^{2}+2\,n-2\,m \right)  \left( z-{z_0} \right) + \frac12\,{
z_0}\,{(z-z_0)}^{2}+\\
\\
+\frac{4}{45}\left({ {3}}-n{{z_0}}^{2}-{n}^{2}-mn+m{{z_0
}}^{2}-{m}^{2}+\frac{3}{2}{z_0}{a_1}-\frac14
{{z_0}}^{4} \right) \left( z-{z_0} \right) ^{3}+\\
\\
+ \frac{1}{18}\left( 2\,m{z_0}-{{z_0}}^{3}-2\,n{z_0}+3\,{a_1}
\right) \left( z-{z_0} \right) ^{4}+{a_6}\, \left( z-{z_0} \right)
^{5}+...
\end{gathered}\end{equation}

Using the series \eqref{7.1} we obtain the following relations for
the zeros of the generalized Hermite polynomials

\begin{equation}\begin{gathered}
\label{7.2}\sum_{k\neq j}^{nm}\frac{1}{(z_k-z_j)}+a_1=0,
\end{gathered}\end{equation}

\begin{equation}\begin{gathered}
\label{7.3}\sum_{k\neq
j}^{nm}\frac{1}{(z_k-z_j)^2}+\frac{z_{j}^{2}}{3}+\frac{2\,(n-m)}{3}=0,
\end{gathered}\end{equation}

\begin{equation}\begin{gathered}
\label{7.4}\sum_{k\neq j}^{nm}\frac{1}{(z_k-z_j)^3}+
\frac{z_j}{2}=0,
\end{gathered}\end{equation}

\begin{equation}\begin{gathered}
\label{7.5}\sum_{k\neq
j}^{nm}\frac{1}{(z_k-z_j)^4}-\frac{z_{j}^{4}}{45}+
\frac{4\,(m-n)z_{j}^{2}}{45}\,+
\frac{2\,a_1\,z_j}{15}\,+\\
\\ +\frac{4}{45}\left({{3}}-{m}^{2}-m\,n-{n}^{2}\right)
=0.
\end{gathered}\end{equation}

The generalized Hermite polynomial $H_{4,4}(z)$ is \cite{Clarkson04}

\begin{equation}\begin{gathered}
\label{7.6}H_{4,4}(z) ={z}^{16}+15\,{z}^{12}-{\frac
{225}{8}}\,{z}^{8}+{\frac {7875}{16}}\,{z}^{4}+{\frac {23625}{256}}.
\end{gathered}\end{equation}

\begin{figure}[h]
\centerline{\epsfig{file=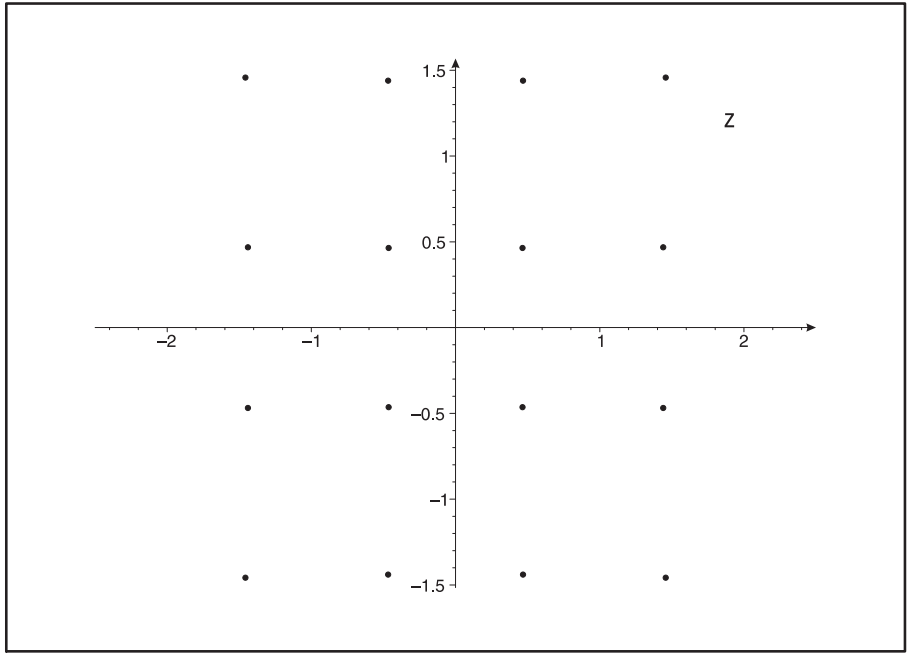,angle=-90,width=120mm}}
\caption{Configuration of zeros of the generalized Hermite
polynomial $H_{4,4}(z)$}\label{Fig5:z_post}
\end{figure}

The configuration of the zeros of the generalized Hermite polynomial
$H_{4,4}(z)$ are illustrated in the Figure 5.

\section {Conclusion}

In this work we have studied special polynomials associated with the
rational solutions of the second, the third, and the fourth
Painlev\'{e} equations. We have shown that if a family of
polynomials satisfies an n-linear differential equation, then much
information concerning the polynomials can be obtained from the
asymptotic study of the solutions of a differential equation related
to the original n-linear one. In particular, it can be found the
degree of each polynomial, symmetric functions of its roots and some
other correlations for the roots called the generalized Stieltjes
relations. Also one can obtain possible degrees of the roots. In
this work we have derived relations for the roots of the following
polynomials: the Laguerre polynomials, the Umemura polynomials, the
Ohyama polynomials, the generalized Okamoto polynomials, and the
generalized Hermite polynomials. Our approach can be applied to
other families of polynomials if there exists a differential
equation satisfied by the family in question \cite{Kudryashov05,
Kudryashov06, Hone01, Cosgrove01}.

\section {Acknowledgments}

This work was supported by the International Science and Technology
Center under Project B 1213.

\end{document}